\pdfoutput=1
\documentclass[aps,aip,twocolumn,superscriptaddress,amsmath]{revtex4-1}
\usepackage{graphicx}
\usepackage{float}

\begin{document}
\title{Electronic structures of [111]-oriented free-standing InAs and InP nanowires}
\author {Gaohua Liao}
\affiliation{Department of Applied Physics, School of Physics and Electronics, Hunan University, Changsha 410082, China}
\affiliation{Key Laboratory for the Physics and Chemistry of Nanodevices and Department of Electronics, Peking University, Beijing 100871, China}
\author {Ning Luo}
\affiliation{Key Laboratory for the Physics and Chemistry of Nanodevices and Department of Electronics, Peking University, Beijing 100871, China}
\author {Ke-Qiu Chen}
\email[Corresponding author. Electronic address: ]{keqiuchen@hnu.edu.cn}
\affiliation{Department of Applied Physics, School of Physics and Electronics, Hunan University, Changsha 410082, China}
\author {H. Q. Xu}
\email[Corresponding author. Electronic address: ]{hqxu@pku.edu.cn}
\affiliation{Key Laboratory for the Physics and Chemistry of Nanodevices and Department of Electronics, Peking University, Beijing 100871, China}
\affiliation{Division of Solid State Physics, Lund University, Box 118, S-221 00 Lund, Sweden}
\date{\today}

\begin{abstract}
We report on a theoretical study of the electronic structures of the [111]-oriented, free-standing, zincblende InAs and InP nanowires with hexagonal cross sections by means of an atomistic $sp^{3}s^{*} $, spin-orbit interaction included, nearest-neighbor, tight-binding method. The band structures and the band state wave functions of these nanowires are calculated and the symmetry properties of the bands and band states are analyzed based on the $C_{3v}$ double point group. It is shown that all bands of these nanowires are doubly degenerate at the $\Gamma$-point and some of these bands will split into non-degenerate bands when the wave vector $k$ moves away from the $\Gamma$-point as a manifestation of spin-splitting due to spin-orbit interaction. It is also shown that the lower conduction bands of these nanowires all show simple parabolic dispersion relations, while the top valence bands show complex dispersion relations and band crossings. The band state wave functions are presented by the spatial probability distributions and it is found that all the band states show $2\pi/3$-rotation symmetric probability distributions. The effects of quantum confinement on the band structures of the [111]-oriented InAs and InP nanowires are also examined and an empirical formula for the description of quantization energies of the lowest conduction band and the highest valence band is presented. The formula can simply be used to estimate the enhancement of the band gaps of the nanowires at different sizes as a result of quantum confinement.
\end{abstract}


\maketitle

\section{INTRODUCTION}
In recent years, InAs and InP nanowires have attracted considerable attention due to their potential application in nanoelectronics and optoelectronics.\cite{Hang-1,Dayeh-1,Nilsson-1,Li-2014,Wang-1,Wallentin-1,Anttu-1,Cui-1,Anttu-2010,Boxberg-2010,Boxberg-2013,Duan-1,Dayeh-2,Joyce-1,Jiang-1,Mohan-1,Pitanti-1} Bulk InAs is a semiconductor with a small direct band gap (0.37 eV), low electron effective mass and high carrier mobility and has been widely used for the development of high-speed electronics. Bulk InP has a large direct band gap (1.34 eV) and excellent optical properties, and has been widely used in high-power electronics and high-performance optoelectronics. Owing to these excellent material characteristics and their intrinsic low dimensionality, InAs and InP nanowires have been explored to realize various novel devices and systems, including nanowire field-effect transistors,\cite{Hang-1,Nilsson-1,Dayeh-1,Li-2014} light emitting devices,\cite{Wang-1,Duan-1}, solar cells,\cite{Wallentin-1,Anttu-1,Cui-1} and superconducting quantum devices.\cite{Ishibashi-2011,Abay-1,Abay-2,Abay-3} It has also been demonstrated that InAs nanowires exhibit strong electron spin-orbit coupling and these nanowires are desired materials for realization of topological superconducting systems in which Majorana bound states can be present and exploited for topological quantum computation.\cite{Kouwenhoven,Deng-2012,Marcus-2013,Deng-2014,Lee-1,Das-1} Epitaxially grown InAs and InP nanowires can have zincblende lattice structures and are most commonly oriented along $\langle111\rangle$ crystallographic directions. Typical diameters of these nanowires are in the range of a few nanometers to about one hundred nanometers.

\par
Previously, the electronic structure have been studied for the InAs and InP nanowires oriented along a $\langle100\rangle$ crystallographic direction.\cite{Xu-7} In this paper, we report on a theoretical study of the electronic structure of InAs and InP nanowires oriented along the [111] crystallographic directions. In comparison with an InAs or InP nanowire oriented along a $\langle100\rangle$ direction, unit cells in a nanowire of similar size oriented along the [111] direction are much larger and the computation for the electronic structure is much more demanding. A $sp^{3}s^{*}$ nearest-neighbor, spin-orbit interaction included, tight-binding formalism is employed in the calculation for the band structures and wave functions of the [111]-oriented InAs and InP nanowires. We show that although the lowest conduction bands of these nanowires display good parabolic dispersions, the top valence bands exhibit complex structures. Furthermore, all the band states at the $\Gamma$-point are doubly degenerate. Some of them will however split into nondegenerate bands when the wave vector $k$ moves away from the $\Gamma$-point. In addition, the wave functions of the band states of the InAs and InP nanowires show characteristic patterns with symmetries as described by the irreducible representations of the $C_{3v}$ double point group and could in general not be reproduced by the calculations based on simple one-band effective mass theory.

\par
The rest paper is organized as follows. In Section II, a brief description for the calculation method is presented. Section III is devoted to the description and discussion of the calculated electronic structures of the [111]-oriented InAs and InP nanowires with hexagonal cross sections of different sizes. Finally, the paper is summarized in Section IV.

\section{Method of Calculations}

\par
Theoretical methods, such as first-principles methods,\cite{Cahangirov-1,Keqiu-1,Keqiu-2,Keqiu-3} $\bf{k}\cdot \bf{p}$ method,\cite{Lassen-1,Kishore-1,Kishore-2} pseudopotential methods,\cite{WangLW-1}, and tight-binding methods \cite{Xu-3,Xu-4,Xu-5,Xu-6,Xu-7,Xu-11,Niquet-1,Niquet-2,Lind-1}, have been employed for the study of semiconductor nanowires. In comparison with other methods, tight-binding methods formed in a $sp^{3}$, $sp^{3}s^{*}$ or $sp^{3}d^{5}s^{*}$ basis \cite{Chadi-1,Vogl-1,Jancu-1,Boykin-1} have been proved to be more powerful for the calculations of the electronic structures of semiconductor nanowires with diameters in the range of a few nanometers to more than 100 nanometers in the whole Brillouin zone.

\par
Here, we employ the $sp^{3}s^{*}$ nearest-neighbor, spin-orbit interaction included, tight-binding formalism in the calculations for the electronic structures of the [111]-oriented InAs and InP zincblend nanowires. In the tight-binding formalism, Bloch sums of the form\cite{Carlo-1,Vogl-1}
\begin{equation}
\label{eq01}
|\alpha,\nu,\bf{k}\rangle=\frac{1}{\sqrt{N}}\sum_{R} e^{i\bf{k}\cdot\bf{R}_{\nu}}|\alpha,\bf{R}_{\nu}\rangle ,
\end{equation}
are used as a basis, where $|\alpha,\bf{R}_{\nu}\rangle$ stands for an atomic orbital $\alpha$ at position $\bf{R}_{\nu}$, $\bf N$ is the number of lattice sites. In the $sp^{3}s^{*}$ nearest-neighbor, spin-orbit interaction included, tight-binding formalism, the atomic orbitals are chosen as 10 localized, spin-dependent orbitals. In the basis of the Bloch sums, the Hamiltonian $\bf{H}$ can be written in a matrix form with matrix elements given by
\begin{equation}
\label{eq02}
\bf{H}_{\alpha\nu,\beta\xi}(k)=\sum_{R}e^{-i\bf{k}\cdot(\bf{R}_{\nu}^{\prime}-\bf{R}_{\xi})}\langle{\alpha,\bf{R}_{\nu}^{\prime}|} \bf{H}|{\beta,\bf{R}_{\xi}}\rangle ,
\end{equation}
and the eigenfunctions in a form of
\begin{equation}
\label{eq03}
{|n,\bf{k}}\rangle=\sum_{\alpha,\nu}c_{n,\alpha\nu}|{\alpha,\nu,\bf{k}}\rangle .
\end{equation}
In the nanowire geometry, only the translational symmetry along the growth direction is preserved and an unit cell in a nanowire with a large lateral size ($>$10 nm) consists of an extremely large number of atoms and the resulting Hamiltonian matrix becomes too large in size to be solved by a standard diagonalization procedure. In this work, the Lanczos algorithm \cite{Golub-1} is employed to solve for the eigenvalues and eigenvectors of Eq.~(\ref{eq02}).

\par
\begin{figure}[t]
\begin{center}
    \includegraphics[width=85mm]{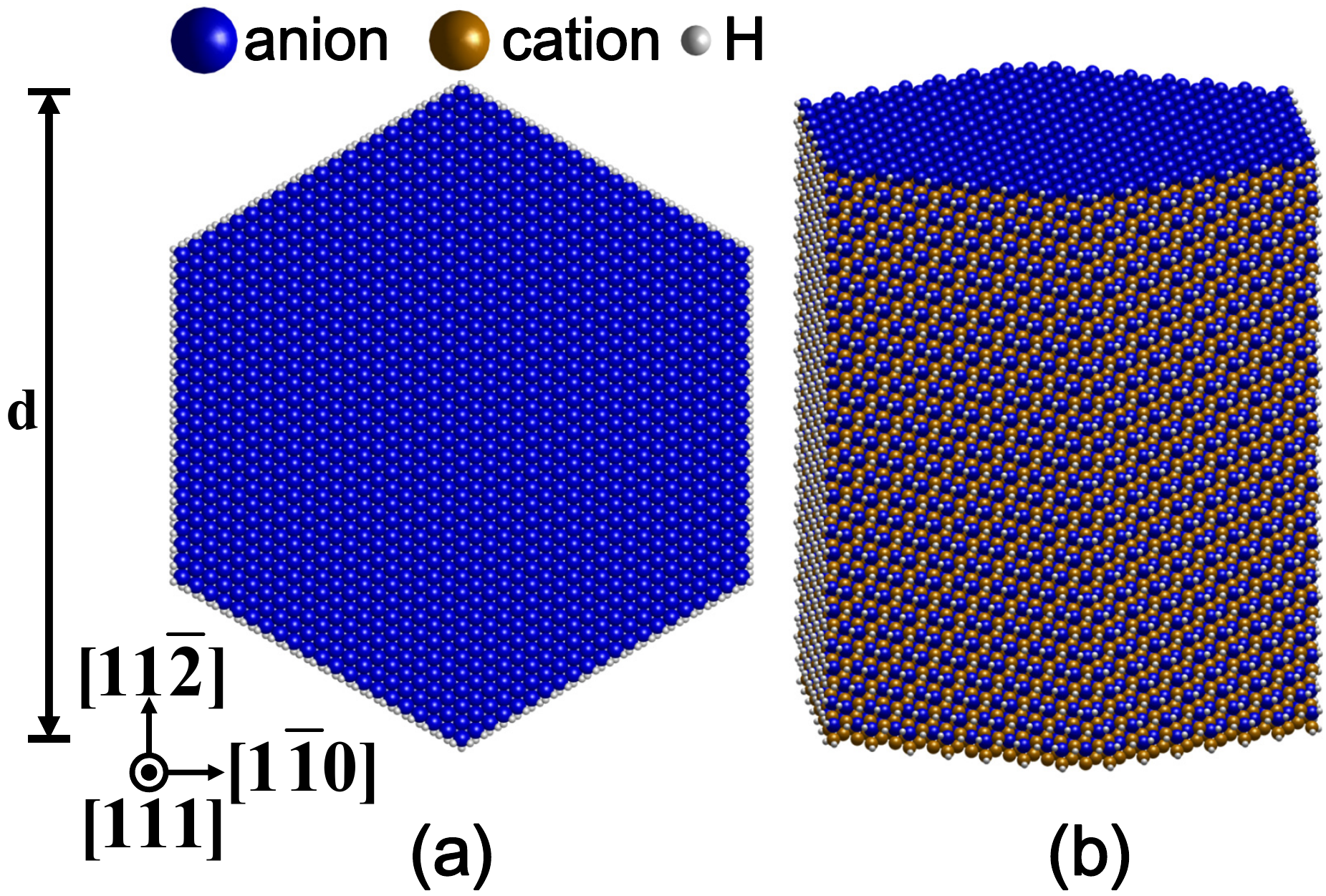}
              \caption{(Color online) Atomistic model structure of a [111]-oriented binary compound nanowire  employed in this work. (a) Top view of the nanowire with size $d$. (b) Side view of a section of the nanowire. The size of the unit cell along the [111] crystallographic direction is $\sqrt{3}a$ where $a$ represents for the lattice constant of the corresponding bulk material. The surface dangling bonds of the nanowires are passivated using hydrogen atoms. Blue, brown and gray spheres represent for anion, cation and hydrogen atoms, respectively.}
        \label{Fig:model}
\end{center}
  \end{figure}

The atomistic model structure of the considered InAs and InP nanowire systems is displayed in Fig.~\ref{Fig:model}. These nanowires are zincblende crystals oriented along the [111] crystallographic direction and  have a hexagonal cross section and $\{1\bar{1}0\}$ facets. The period of unit cells in a nanowire is $\sqrt{3} a$ and the Brillouin zone is defined as $-2\pi/(\sqrt{3}a) \leq k \leq 2\pi/(\sqrt{3} a)$, where $a$ is the lattice constant of the corresponding bulk material. The lateral size of these nanowires is defined as the distance $d$ between two most remote corners in the hexagonal cross section as shown in Fig.~\ref{Fig:model}(a). In an atomistic model, $d$ takes discrete values of $d=2na/\sqrt{6}$, where $n$ is a positive integer number. The dangling bonds at the surface of the nanowires are passivated using hydrogen atoms to eliminate the effects of these dangling bonds on the band states near the fundamental band gaps. In the calculations, the InAs and InP tight-binding  parameters are taken from Ref.~\onlinecite{Klimech-1} and the parameters involved the passivation hydrogen atoms are determined by the procedure presented in Refs.~\onlinecite{Xu-1} and~\onlinecite{Xu-2}. For further details about the method of calculations, we refer to Refs.~\onlinecite{Xu-6} and ~\onlinecite{Xu-11}.

\begin{figure*}[t]
\begin{center}
    \includegraphics[width=170mm]{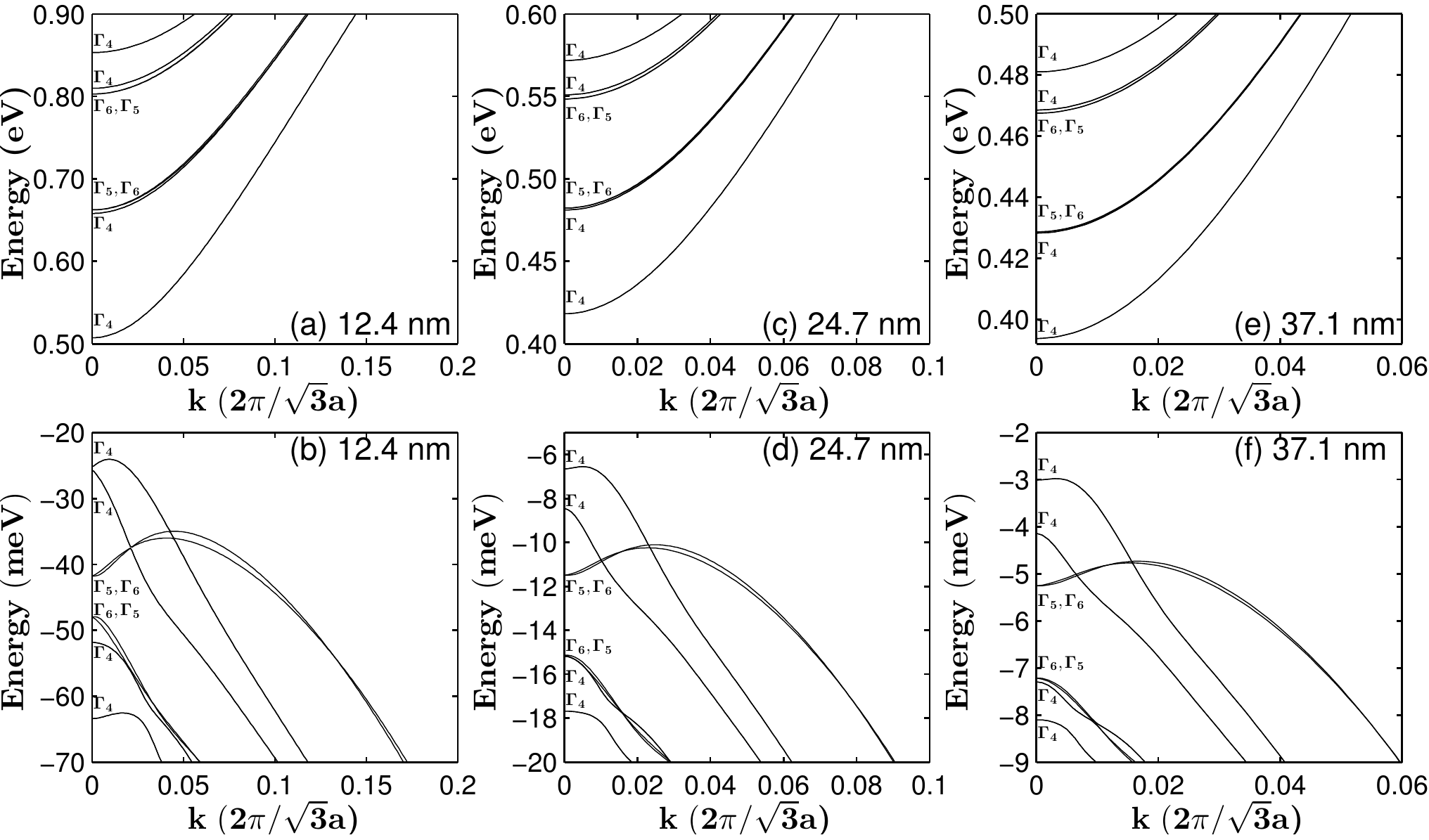}
              \caption{Band structures of the [111]-oriented InAs nanowires with hexagonal cross sections of  lateral sizes $d=12.4$ nm [(a) and (b)], $d=24.7$ nm [(c) and (d)], and $d=37.1$ nm [(e) and (f)]. The bands are labeled according to the irreducible representations, $\Gamma_{4}$, $\Gamma_{5}$ and $\Gamma_{6}$, of the $C_{3v}$ double point group. Here we note that all bands are doubly degenerate at $k=0$ (the $\Gamma$-point). In the figure, labels $\Gamma_{5}$ and $\Gamma_{6}$ are ordered in such a way that the first one labels a band that initially has the lower energy after splitting as the wave vector moves away from the $\Gamma$-point.}
        \label{Fig:[111]InAsband}
 \end{center}
  \end{figure*}

\section{results and discussion}
\par
In this section, we present the results of calculations for the band structures and wave functions of the [111]-oriented InAs and InP nanowires with hexagonal cross sections. These nanowires are symmetric under the operations of the $C_{3v}$  point group (with its rotational axis along the nanowire axis). The corresponding double point group has one two-dimensional irreducible representation $\Gamma_{4}$ and two one-dimensional irreducible representations $\Gamma_{5}$ and $\Gamma_{6}$.\cite{Tinkham-1,Melvin-1,Yu-1} However, at the $\Gamma$ point, all the band states are doubly degenerate due to the Kramers' degeneracy. When the wave vector $k$ moves away form the $\Gamma$ point, the band structure Hamiltonian does not possess time-reversal symmetry and the $\Gamma_{5}$ and $\Gamma_{6}$ bands would split into nondegenerate bands. However, the $\Gamma_{4}$ bands will remain doubly degenerate at all wave vector points.

\subsection{Band structures}

  \par
\begin{figure*}
 \centering
    \includegraphics[width=170mm]{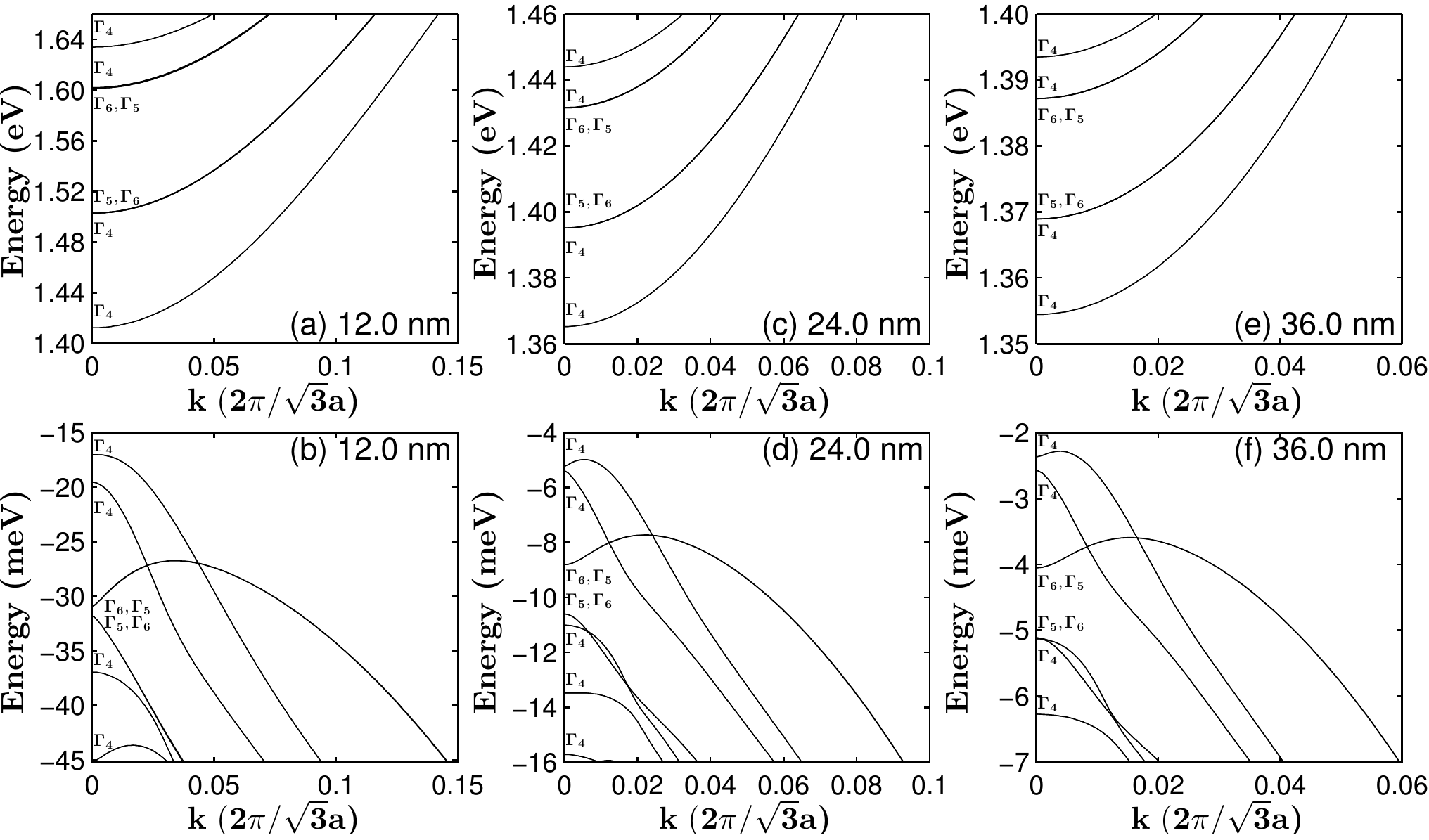}
       \caption{Band structures of the [111]-oriented InP nanowires with hexagonal cross sections of lateral size $d=$12.0 nm [(a) and (b)], $d=$ 24.0 nm [(c) and (d)], and $d=$ 36.0 nm [(e) and (f)]. The bands are labeled according to the irreducible representations, $\Gamma_{4}$, $\Gamma_{5}$ and $\Gamma_{6}$, of the $C_{3v}$ double point group. Here we note that all bands are double degenerate at $k=0$ (the $\Gamma$-point). In the figure, labels $\Gamma_{5}$ and $\Gamma_{6}$ are ordered in such a way that the first one labels a band that initially has the lower energy after splitting as the wave vector moves away from the $\Gamma$ point.}
        \label{Fig:[111]InPband}
  \end{figure*}

We first present the calculated band structures of the [111]-oriented InAs and InP nanowires. Figure~\ref{Fig:[111]InAsband} shows the calculated band structures of the [111]-oriented InAs nanowires with hexagonal cross sections of size $d=$ 12.4, 24.7 and 37.1 nm (i.e., $n=$ 25, 50 and 75) and Fig.~\ref{Fig:[111]InPband} shows the calculated band structures of the [111]-oriented InP nanowires with hexagonal cross sections of size $d=$ 12.0, 24.0 and 36.0 nm (i.e., $n=$ 25, 50 and 75), respectively.  In general, the band gaps of the [111]-oriented InAs and InP nanowires are both increased as the lateral sizes $d$ are decreased. It is also found that the low-energy conduction bands of the [111]-oriented InAs and InP nanowires show simple, parabolic dispersions around the $\Gamma$-point and the top valence bands of the [111]-oriented InAs and InP nanowires show complex structures. The simple, parabolic dispersion characteristics seen in these low-energy conduction bands have also been found in the calculations for the band structures of the [100]-oriented InAs and InP nanowires with  square cross sections.\cite{Xu-7} However, the complex structures seen in the top valence bands of the [111]-oriented InAs and InP nanowires are distinctively different from the top valence bands of these [100]-oriented InAs and InP nanowires.\cite{Xu-7} The former are dominantly characterized by band crossings, while the latter show band anti-crossings. Also, the double maximum structure seen in the topmost valence band of a [100]-oriented InAs or InP nanowire with a square cross section does not fully develop in the [111]-oriented InAs and InP nanowires.

\par
In details, it is seen in Figs.~\ref{Fig:[111]InAsband}(a),~\ref{Fig:[111]InAsband}(c) and \ref{Fig:[111]InAsband}(e) that the lowest and the second lowest conduction bands of a [111]-oriented InAs nanowire with a hexagonal cross section are all $\Gamma_{4}$  symmetric and thus doubly degenerate, and the next two lowest conduction bands are $\Gamma_{5}$  and $\Gamma_{6}$ bands, which are in general non-degenerate bands except for at the $\Gamma$-point. However, it is seen that the second lowest conduction band and the $\Gamma_{5}$  and $\Gamma_{6}$  bands are very close in energy and they tend to the formation of a nearly four-fold degenerate band as the nanowire lateral size is increased. The next two lowest conduction bands are again $\Gamma_{5}$  and $\Gamma_{6}$ bands. These two bands are close in energy to the next lowest $\Gamma_{4}$  band to form another nearly four-fold degenerate band.  Figures~\ref{Fig:[111]InAsband}(b),~\ref{Fig:[111]InAsband}(d) and \ref{Fig:[111]InAsband}(f) show that the two highest valence bands of the [111]-oriented InAs nanowire with a hexagonal cross section are $\Gamma_{4}$  symmetric, doubly degenerate bands. The next two highest valence bands are a $\Gamma_{5}$ and a $\Gamma_{6}$ band whose states are, in general, non-degenerate, but very close in energy,  except for at the $\Gamma$ point at which the two band states are exactly degenerate due to the presence of time reversal symmetry in the Hamiltonian.
The next two highest valence bands of the [111]-oriented nanowires are again the $\Gamma_{5}$ and $\Gamma_{6}$ bands with degenerate or nearly degenerate energies. At a large nanowire lateral size (see the cases for $d=$24.7 and 37.1 nm), these two bands are very close to the next highest $\Gamma_{4}$ valence band in energy, leading to the formation of a nearly four-fold nearly degenerate valence band.

\par
Figure~\ref{Fig:[111]InPband} shows the band structures of the [111]-oriented InP nanowires with hexagonal cross sections of lateral size $d=$12.0, 24.0 and 36.0 nm. The band structures exhibit similar characteristics as the [111]-oriented InAs nanowires. In particular, all the low-energy conduction band show simple parabolic dispersion structures and the top valence bands show complex dispersive, band-crossing characteristics. It is also seen that the symmetry ordering of the bands at the $\Gamma$-point remain the same as in the [111]-oriented InAs nanowires. However, some differences can be identified. In particular,  as can be seen from Fig.~\ref{Fig:[111]InPband} that the second lowest conduction band ($\Gamma_4$ band) and the third and fourth lowest conduction bands ($\Gamma_5$ and $\Gamma_6$ bands) of the [111]-oriented InP nanowires become hardly distinguishable, forming a nearly four-fold degenerate conduction band, even at a small lateral size. The next lowest $\Gamma_4$, $\Gamma_5$ and $\Gamma_6$ conduction bands are also hardly distinguishable in energy and form a nearly four-fold degenerate conduction band already at a small nanowire lateral size. Overall, we see in Fig.~\ref{Fig:[111]InPband} that all the $\Gamma_5$ and $\Gamma_6$ bands in the [111]-oriented InP nanowires are paired up to form nearly doubly degenerate bands.

\begin{table}[t]
\caption{Parameters $p_{1}$, $p_{2}$, and $p_{3}$ in Eq.~(\ref{eq04}) obtained by fitting the equation to the calculated energies at the $\Gamma$ point of the lowest conduction band and the highest valence band of the [111]-oriented InAs and InP nanowires with hexagonal cross sections. }
\begin{center}
\begin{tabular*}{85mm}{@{\extracolsep{\fill}}c c c c c  }
\hline\hline
Material &  Band shift & $p_{1}$             &$p_{2}$             &$p_{3}$     \\
                & (eV)    &(eV$^{-1}$nm$^{-2}$) &(eV$^{-1}$nm$^{-1}$) &(eV$^{-1}$) \\
\hline
InAs                 &$\Delta E_{c}$  &0.01901  &0.32183  &0.31233 \\
\qquad           &$\Delta E_{v}$  &-0.18486  &-1.26185  &-0.17868 \\
InP          &$\Delta E_{c}$  &0.07651  &0.29013  &0.75008 \\
\qquad          &$\Delta E_{v}$  &-0.33648  &-1.89510  &-0.11117 \\
\hline\hline
\end{tabular*}
\end{center}
\label{tab:Table1}
\end{table}

\begin{figure}[t]
\begin{center}
\includegraphics[width=85mm]{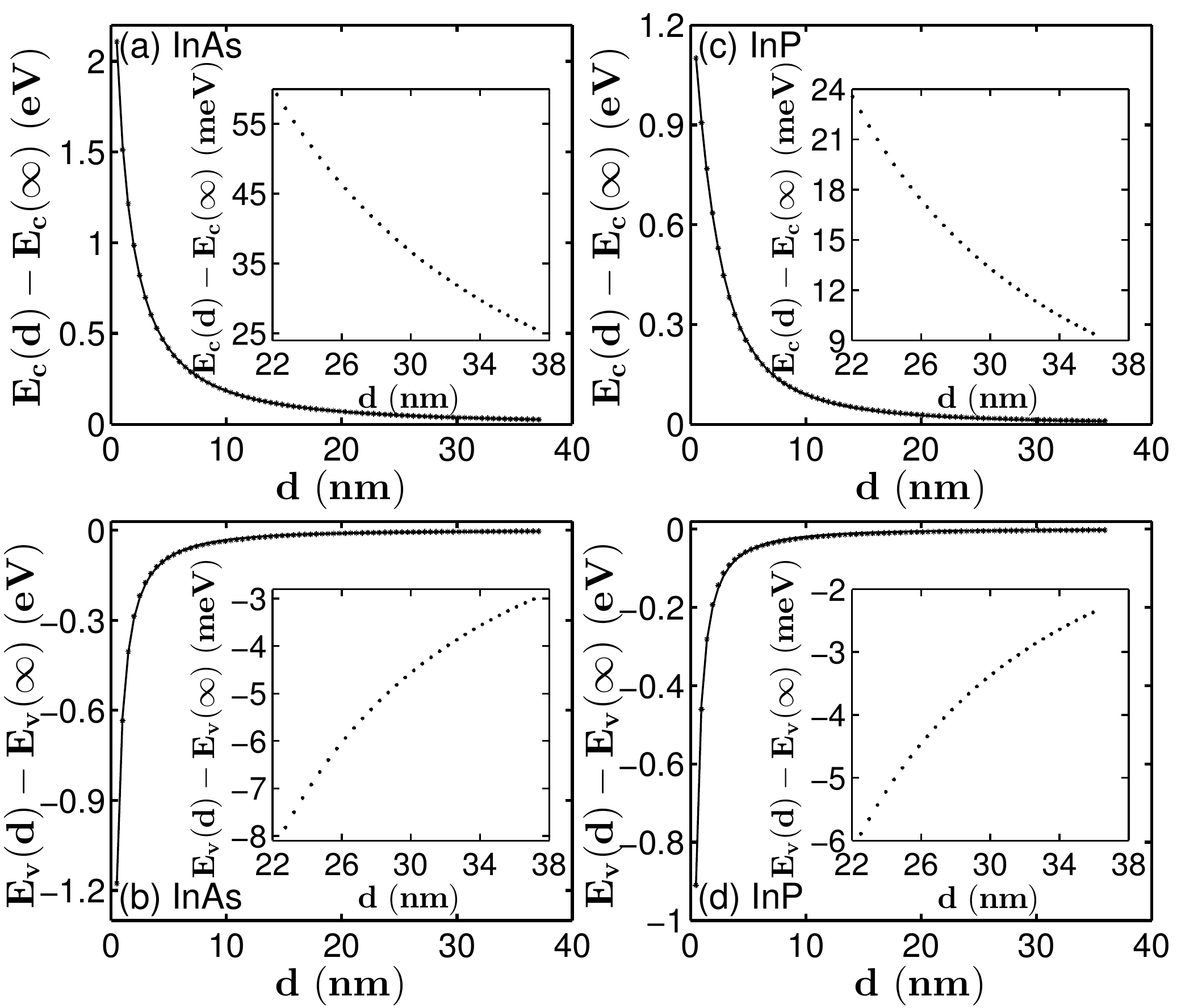}
\caption{Lowest conduction band electron and highest valence band hole confinement energies in the [111]-oriented InAs and InP nanowires with hexagonal cross sections as functions of cross section size $d$. Panels (a) and (b) show the results for the InAs nanowire, while panels (c) and (d) show the results for the InP nanowire. The calculated data are presented by symbols ``$*$" and the solid lines are the results of fittings based on Eq.~(\ref{eq04}) with the fitting parameters listed in Table~\ref{tab:Table1}. The insets show the zoom-in plots of the calculated confinement energies in the nanowires at large sizes.}
\label{Fig:[111]InAsInPcon}
\end{center}
\end{figure}

\par
As we mentioned above, the conduction bands and the valence bands move apart in energy as the cross section sizes of the [111]-oriented InAs and InP nanowires are decreased as a result of quantum confinement. It is found that the calculated edge energies of the lowest conduction bands and the topmost valence bands, $E_{c}(d)$ and $E_{v}(d)$, of the [111]-oriented InAs and InP nanowires with different lateral sizes $d$ can be fitted to the following expression,
\begin{equation}
\label{eq04}
\Delta E_\alpha(d) =E_\alpha(d)-E_\alpha (\infty)=\frac{1}{p_{1} d^{2}+p_{2} d+p_{3}},
\end{equation}
where $\alpha=c$ or $v$, $E_v(\infty)$ and $E_c(\infty)$ are the energies at the bottom of the conduction band and the top of valence band of the bulk materials, $\Delta_c$ and $\Delta_v$ stand for the energy shifts in the bottom conduction band and the top valence band due to quantum confinement, and $p_1$, $p_2$ and $p_3$ are fitting parameters. The results of fitting are shown in Fig.~\ref{Fig:[111]InAsInPcon} with the obtained fitting parameters listed in Table~\ref{tab:Table1}. It is seen that the effect of quantum confinement on the conduction band of the InAs nanowires is stronger than that of InP nanowires,  consistent with the fact of InAs has a smaller electron effective mass. For example, the quantum confinement energy of electrons at the conduction band edge of the [111]-oriented InAs nanowire with a hexagonal cross section of size $d\sim$ 22 nm is $\sim$ 60 meV and the corresponding quantum confinement energy in the [111]-oriented InP nanowire of the same lateral size is $\sim$ 24 meV. When the InAs nanowire has a lateral size of $d\sim$ 38 nm, the electron quantization energy at the conduction band edge can still be $\sim 24$ meV. In comparison, for the InP nanowire with size $d\sim$ 38 nm, the electron quantization energy at the conduction band edge is only $\sim 8$ meV. The quantum confinement energies of holes in the valence bands are, in general, very small when compared with the corresponding quantum confinement energies of electrons at the conduction bands of the nanowires. For example, the quantum confinement energies are only $\sim$ 8 meV and $\sim$ 6 meV at the valence bands of the InAs and InP nanowires with a lateral size $d\sim$ 22 nm, respectively.  As the lateral size $d$ is increased to $\sim$ 38 nm, the quantum confinement energies of holes in the valence bands of InAs and InP nanowires are decreased to $\sim$ 3 meV and $\sim$ 2 meV, respectively.

\subsection{Wave functions}

\par
\begin{figure}[t]
\begin{center}
\includegraphics[width=85mm]{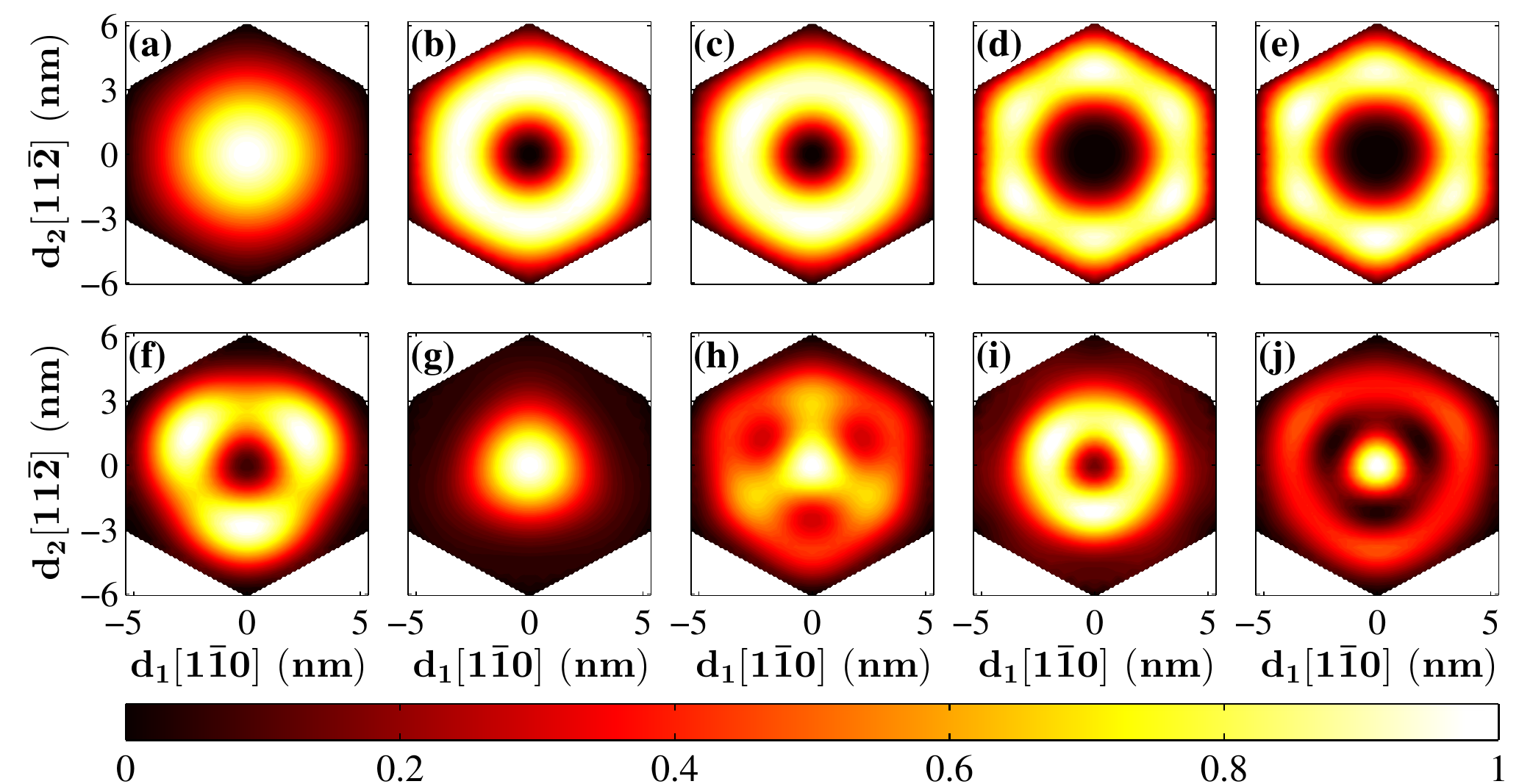}
\caption{(Color online) Wave functions of the five lowest conduction band states and the five highest valence band states at the $\Gamma$-point of the [111]-oriented InAs nanowire with a hexagonal cross section of size $d=12.4$ nm. The wave functions are presented by the probability distributions on a (111)-plane of In atoms whose value at each atomic site is calculated by summing up the squared amplitudes of all the atomic orbital components on the atomic site and is normalized within each panel by the highest value found in the panel. Panels (a) to (e) show the wave functions of the five lowest conduction band states at the $\Gamma$-point, while panels (f) to (j) show the wave functions of the five highest valence band states at the $\Gamma$-point.}
\label{InAsN25wave}
\end{center}
\end{figure}
\begin{figure}[t]
\begin{center}
\includegraphics[width=85mm]{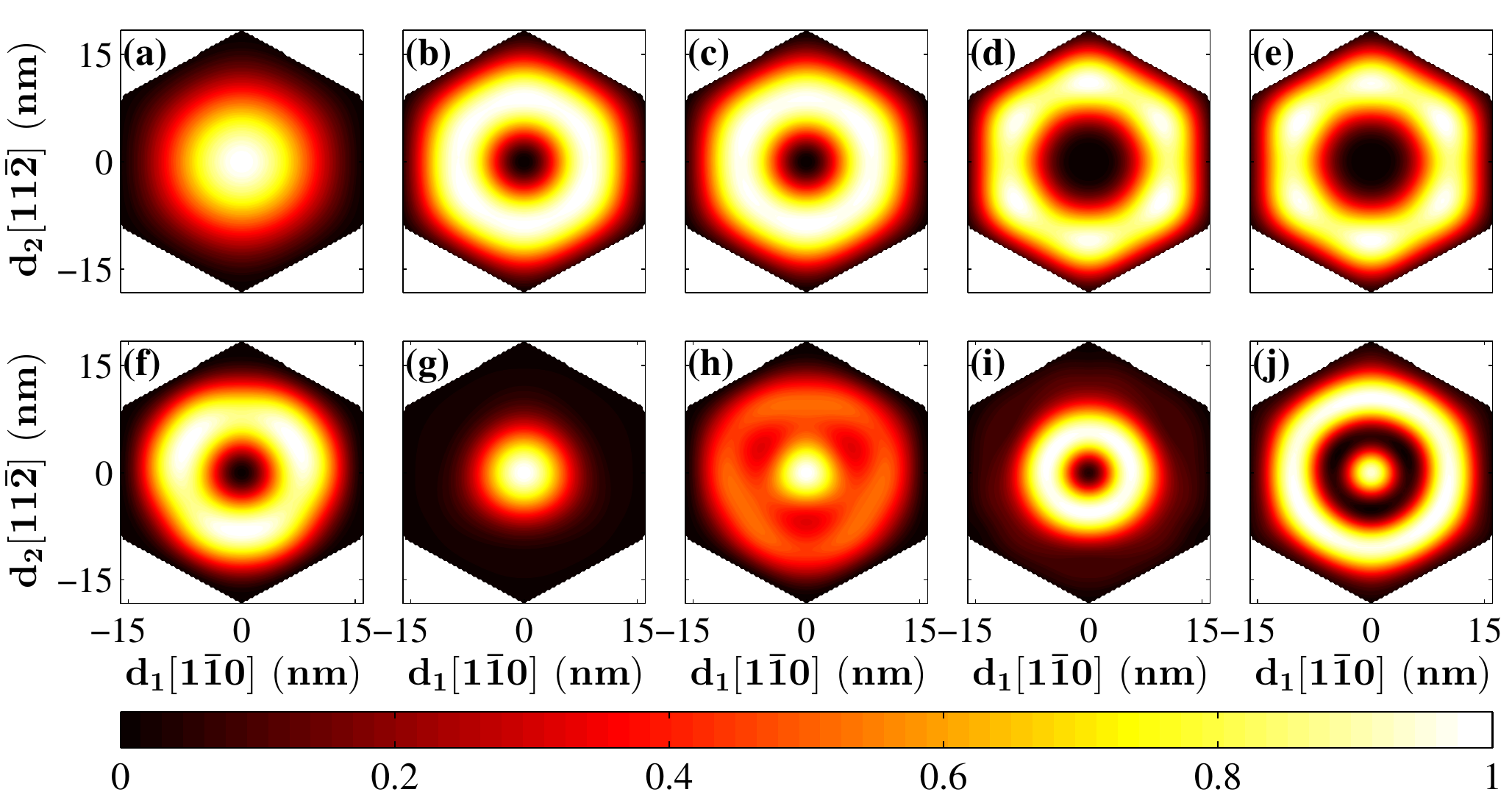}
\caption{(Color online) The same as in Fig.~\ref{InAsN25wave} but for the [111]-oriented InAs nanowire with a hexagonal cross section of size $d=37.1$ nm.}
 \label{InAsN75wave}
 \end{center}
\end{figure}
\begin{figure}[!t]
\begin{center}
\includegraphics[width=85mm]{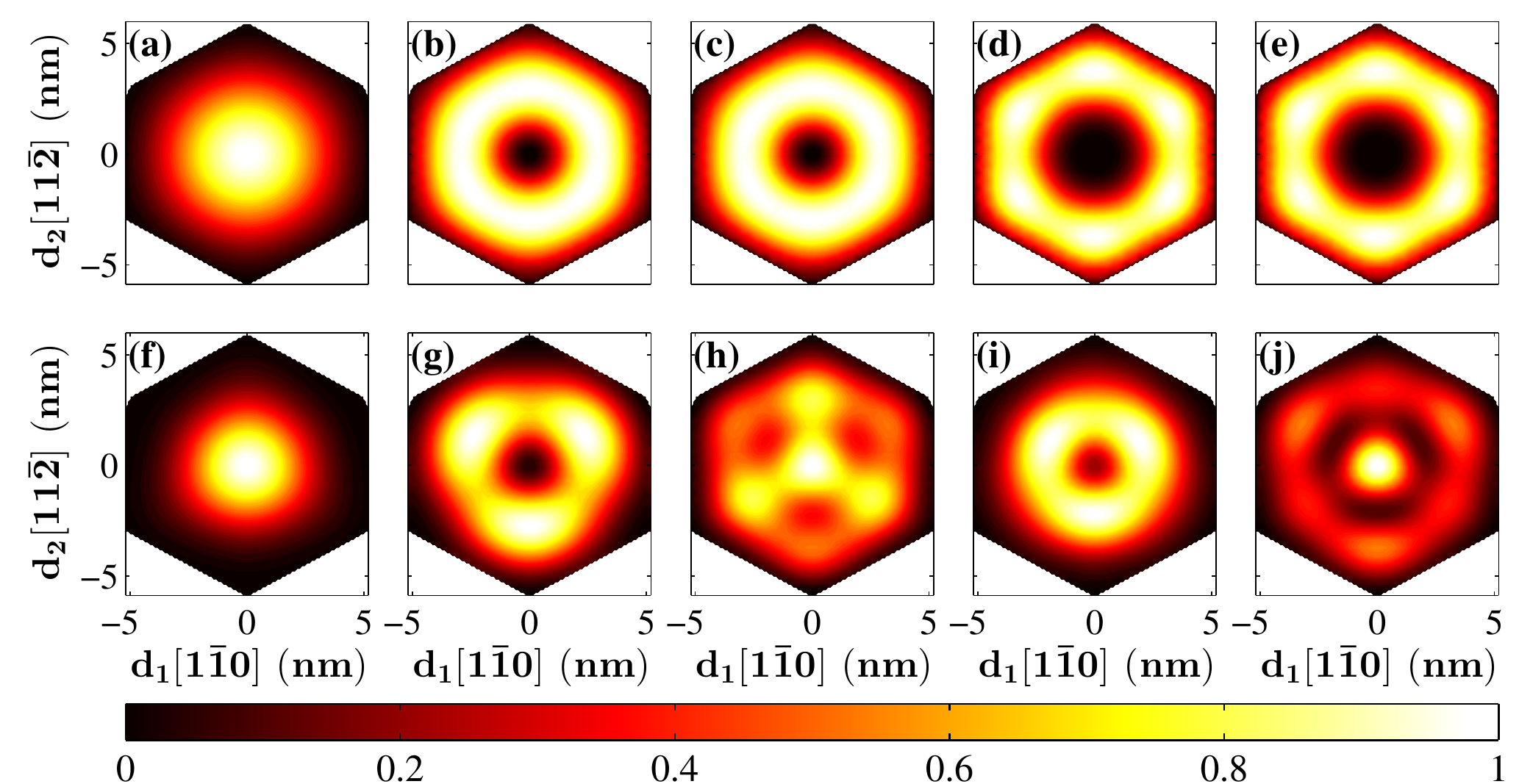}
\caption{(Color online) The same as in Fig.~\ref{InAsN25wave} but for the [111]-oriented InP nanowire with a hexagonal cross section of size $d=12.0$ nm.}
\label{InPN25wave}
\end{center}
\end{figure}
\begin{figure}[t]
\begin{center}
\includegraphics[width=85mm]{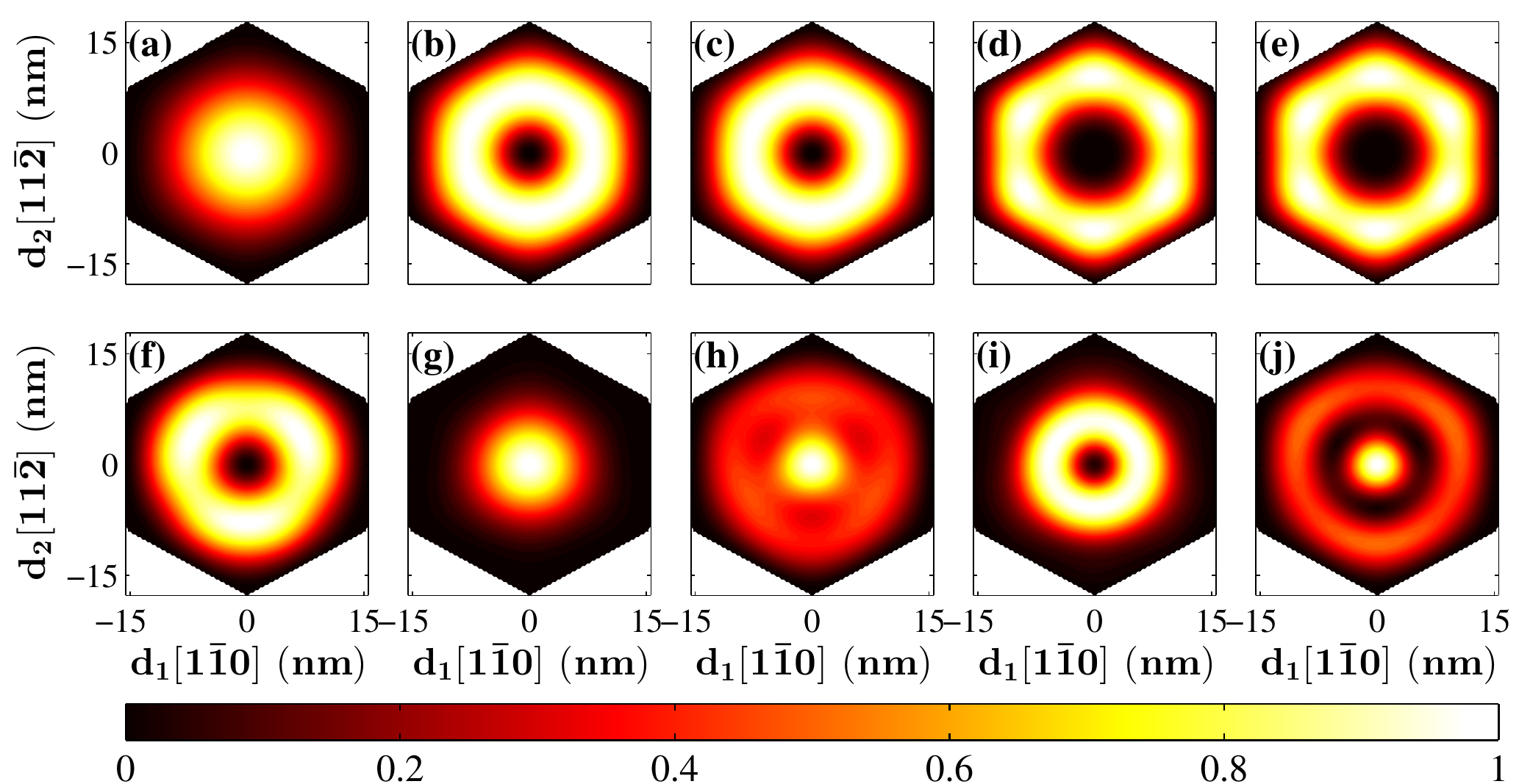}
\caption{(Color online) The same as in Fig.~\ref{InAsN25wave} but for the [111]-oriented InP nanowire with a hexagonal cross section of size $d=36.0$ nm.}
 \label{InPN75wave}
 \end{center}
\end{figure}

The wave functions of band states have also been calculated and analyzed for the [111]-oriented free-standing InAs and InP nanowires.  Figures~\ref{InAsN25wave} to \ref{InPN75wave} show the representative results of the calculations for a few lowest conduction band states and a few highest valence band states of the nanowires of different sizes at the $\Gamma$ point. Here, a wave function is represented by the probability distribution on a (111)-plane of cation (In or Ga) atoms with the probability at each atomic site calculated by summing up the squared amplitudes of all the atomic orbital components on that site and scaled against the maximum value within each graph. As we mentioned above, at the $\Gamma$ point, all the states are doubly degenerate. In Figs.~\ref{InAsN25wave} to \ref{InPN75wave}, we only plot the probability distribution for one of the two spin-degenerate states, since the other one has an identical spacial probability distribution.

\par
Figures~\ref{InAsN25wave} and~\ref{InAsN75wave} show the calculated wave functions for the five lowest conduction band states and the five highest valence band states at the $\Gamma$-point of a [111]-oriented InAs nanowire with a hexagonal cross section of size $d=12.4$ nm ($n$=25) and $d=37.1$  nm ($n$=75), respectively. It is seen that the corresponding conduction band states in the two InAs nanowires with very different sizes have the same special probability distribution characteristics. The lowest conduction band state (a $\Gamma_{4}$-symmetric state) in each InAs nanowire shows a highly symmetric, $s$-like wave function probability distribution and is very localized to the center region of the nanowire. The other four lowest conduction band states in each nanowire all show donut-shaped probability distributions. These four conduction band states can be grouped into two groups based on their special localizations. The second lowest conduction band state (a $\Gamma_{4}$-symmetric  state) and the third lowest conduction band state (a $\Gamma_{5}$- or a $\Gamma_{6}$-symmetric state) are in one group and their wave functions are more localized around the center of the nanowire. The fourth lowest conduction band state (again a $\Gamma_{5}$- or a $\Gamma_{6}$-symmetric state) and the fifth lowest conduction band state (a $\Gamma_{4}$-symmetric state) comprise the second group and their wave functions are less localized to the center region of the nanowire and are clearly of a hexagonal shape with six high probability regions localized at the six hexagonal corners.

\par
The wave functions of the valence band states at the $\Gamma$-point of the two InAs nanowires with different sizes also show similar spacial distribution characteristics. Nevertheless, the distribution patterns of these valence band states are more complex than their conduction band counterparts. Furthermore, it is interesting to see that the probability distribution of the highest valence band state (a $\Gamma_{4}$-symmetric state) of a [111]-oriented InAs nanowire is shaped as a donut with a $2\pi/3$-rotational symmetry, while the second highest valence band state (also a $\Gamma_{4}$-symmetric state) is a strongly localized, $s$-like state. The third highest valence band state (a $\Gamma_{5}$- or a $\Gamma_{6}$-symmetric state) is also a state more localized inside the nanowire with its distribution pattern consisting of a small triangular region of high probability distribution in the center and a region of low but complex probability distribution close to the surface of the nanowire. The fourth highest valence band state (again a $\Gamma_{5}$-  or a $\Gamma_{6}$-symmetric state) exhibits a donut-shaped probability distribution and is much more localized inside the nanowire when compared with the second-fifth lowest conduction band states. The fifth highest valence band state (a $\Gamma_{4}$-symmetric state) exhibits a pattern of probability distribution consisting of a highly symmetric region strongly localized in the center of the nanowire and a ring-like region localized in the middle between the center and the surface of the nanowire.

\par
Figures~\ref{InPN25wave} and \ref{InPN75wave} show the calculated wave functions for the five lowest conduction band states and the five highest valence band states at the $\Gamma$-point of the [111]-oriented InP nanowires with size $d=$12.0 nm ($n$=25) and size $d=$36.0 nm ($n$=75), respectively. It is seen that the wave functions of the five lowest conduction band states of the InP nanowires have the same spatial probability distribution characteristics as the InAs nanowires. The same is true for the wave functions of the five highest valence band states of the InP nanowires, except for a minor difference, i.e., the ordering of the probability distribution patterns of the two highest valence band states (two $\Gamma_4$-symmetric states) at the $\Gamma$-point of the InP nanowire with the smaller size $d$=12.0 nm is reversed when compared to the InAs nanowire with size $d$=12.4 nm.

\section{Conclusions}

\par
In summary, we present a theoretical study of the electronic structures of the [111]-oriented, free-standing, zincblende InAs and InP nanowires with hexagonal cross sections by means of the atomistic $sp^{3}s^{*} $, spin-orbit interaction included, nearest-neighbor tight-binding method. The band structures and the band state wave functions of these nanowires are calculated and the symmetry properties of the bands and band states are analyzed based on the $C_{3v}$ double point group. It is shown that all bands of these nanowires are doubly degenerate at the $\Gamma$-point and some of these bands will split into non-degenerate bands when the wave vector $k$ moves away from the $\Gamma$-point as a manifestation of spin-splitting due to spin-orbit interaction. It is also shown that the lower conduction bands of these nanowires all show simple parabolic dispersion relations, while the top valence bands show complex dispersion relations and band crossings. The appearance of the band crossings in the top valence bands of the [111]-oriented InAs and InP nanowires is in strong contrast to the band structures of the [100]-oriented InAs and InP nanowires with square cross sections in which the top valence bands are seen to go for anti-crossings.  The wave functions of the band states of these [111]-oriented InAs and InP nanowires are presented by probability distributions on cross sections. It is found that all the band states show $2\pi/3$-rotation symmetric probability distributions and two degenerate band states at the $\Gamma$-point show identical spatial probability distributions. Finally, the effects of quantum confinement on the band structures of the [111]-oriented InAs and InP nanowires are examined and an empirical formula for the description of quantization energies of the lowest conduction band and the highest valence band is presented. The formula can simply be used to estimate the enhancement of the band gaps of the nanowires at different sizes as a result of quantum confinement. We believe that the results presented in this work will provide important information about the electronic structures of the [111]-oriented InAs and InP nanowires and useful guidance for the use of these nanowires in novel nanoelectronic, optoelectronic and quantum devices.

\section*{Acknowledgments}
This work was supported by the National Basic Research Program of China
(Grants No.~2012CB932703 and No.~2012CB932700) and the National Natural Science Foundation of China (Grants No.~91221202, No.~91421303, and No.~61321001).

\newpage


\begin{thebibliography}{}
\bibitem{Hang-1}Q. L. Hang, F. D. Wang, P. D. Carpenter, D. Zemlyanov, D. Zakharov, E. A. Stach, W. E. Buhro, and D. B. Janes, Nano Lett. {\bf8}, 49 (2008).
\bibitem{Dayeh-1}S. A. Dayeh, D. P. R. Aplin, X. T. Zhou, P. K. L. Yu, E. T. Yu, and D. L. Wang, Small {\bf3}, 326 (2007).
\bibitem{Nilsson-1}H. A. Nilsson, P. Caroff, C. Thelander, E. Lind, O. Karlstr\"{o}m, and L.-E. Wernersson, Appl. Phys. Lett. {\bf96}, 153505 (2010).
\bibitem{Li-2014} Q. Li, S. Huang, D. Pan, J. Wang, J. Zhao, and H. Q. Xu, Appl. Phys. Lett. {\bf 105}, 113106 (2014).
\bibitem{Wang-1}J. F. Wang, M. S. Gudiksen, X. F. Duan, Y. Cui, and C. M. Lieber, Science {\bf293}, 1455 (2001).
\bibitem{Duan-1}X. Duan, Y. Huang, Y. Cui, J. Wang, and C. M. Lieber, Nature {\bf409}, 66 (2001).
\bibitem{Wallentin-1}J. Wallentin, N. Anttu, D. Asoli, M. Huffman, I. \r{A}berg, M. H. Magnusson, G. Siefer, P. Fuss-Kailuweit, F. Dimroth, B. Witzigmann, H. Q. Xu, L. Samuelson, K. Deppert, and M. T. Borgstr\"{o}m, Science {\bf339}, 1057 (2013).
\bibitem{Anttu-1}N. Anttu, A. Abrand, D. Asoli, M. Heurlin, I. \r{A}berg, L. Samuelson, and M. Borgstr\"{o}m, Nano Res. {\bf7}, 816 (2014).
\bibitem{Cui-1}Y. Cui, J. Wang, S. R. Plissard, A. Cavalli, T. T. Vu, R. P. van Veldhoven, L. Gao, M. Trainor, M. A. Verheijen, J. E. Haverkort, and E. P. Bakkers, Nano Lett. {\bf13}, 4113 (2013).
\bibitem{Anttu-2010} N. Anttu and H. Q. Xu, J. Nanosci. Nanotechnol. {\bf 10}, 7183 (2010).
\bibitem{Boxberg-2010} F. Boxberg, N. S{\o}ndergaard, and H. Q. Xu, Nano Lett. {\bf 10}, 1108 (2010).
\bibitem{Boxberg-2013} F. Boxberg, N. S{\o}ndergaard, and H. Q. Xu, Adv. Mater. {\bf 24}, 4692 (2012).
\bibitem{Dayeh-2}S. A. Dayeh, D. P. Aplin, X. Zhou, P. K. Yu, E. T. Yu, and D. Wang, Small {\bf3}, 326 (2007).
\bibitem{Joyce-1}H. J. Joyce, J. Wong-Leung, C. K. Yong, C. J. Docherty, S. Paiman, Q. Gao, H. H. Tan, C. Jagadish, J. Lloyd-Hughes, L. M. Herz, and M. B. Johnston, Nano Lett. {\bf12}, 5325 (2012).
\bibitem{Jiang-1}X. Jiang, Q. Xiong, S. Nam, F. Qian, Y. Li, and C. M. Lieber, Nano Lett. {\bf7}, 3214 (2007).
\bibitem{Mohan-1}P. Mohan, J. Motohisa, and T. Fukui, Appl. Phys. Lett. {\bf88}, 133105 (2006).
\bibitem{Pitanti-1}A. Pitanti, D. Ercolani, L. Sorba, S. Roddaro, F. Beltram, L. Nasi, G. Salviati, and A. Tredicucci, Phys. Rev. X {\bf1}, 011006 (2011).
\bibitem{Ishibashi-2011} T. Nishio, T. Kozakai, S. Amaha, M. Larsson, H. A. Nilsson, H. Q. Xu, G. Zhang, K. Tateno, H. Takayanagi, and K. Ishibashi, Nanotechnol. {\bf 22}, 445701 (2011).
\bibitem{Abay-1} S. Abay, H. Nilsson, F. Wu, H. Q. Xu, C. M. Wilson, and P. Delsing, Nano Lett. {\bf 12}, 5622 (2012).
\bibitem{Abay-2} S. Abay, D. Persson, H. Nilsson, H. Q. Xu, M. Fogelstr\"{o}m, V. Shumeiko, and P. Delsing, Nano Lett. {\bf 13}, 3614 (2013).
\bibitem{Abay-3} S. Abay, D. Persson, H. Nilsson, F. Wu, H. Q. Xu, M. Fogelstr\"{o}m, V. Shumeiko, and P. Delsing, Phys. Rev. B {\bf 89}, 214508 (2014).
\bibitem{Kouwenhoven} V. Mourik, K. Zuo, S. M. Frolov, S. R. Plissard, E. Bakkers, and L. P. Kouwenhoven, Science {\bf 336}, 1003 (2012).
\bibitem{Deng-2012} M. T. Deng, C. L. Yu, G. Y. Huang, M. Larsson, P. Caroff, and H. Q. Xu, Nano Lett. {\bf 12}, 6414 (2012).
\bibitem{Marcus-2013} H. O. H. Churchill, V. Fatemi, K. Grove-Rasmussen, M. T. Deng, P. Caroff, H. Q. Xu, and C. M. Marcus, Phys. Rev. B {\bf 87}, 241401(R) (2013).
\bibitem{Deng-2014} M. T. Deng, C. L. Yu, G. Y. Huang, M. Larsson, P. Caroff, and H. Q. Xu, Sci. Rep. {\bf 4}, 7261 (2014).
\bibitem{Lee-1}E. J. H. Lee, X. C. Jiang, R. Aguado, G. Katsaros, C. M. Lieber, and S. De Franceschi, Phys. Rev. Lett. {\bf109} (2012).
\bibitem{Das-1}A. Das, Y. Ronen, Y. Most, Y. Oreg, M. Heiblum, and H. Shtrikman, Nature Phys. {\bf8}, 887 (2012).
\bibitem{Cahangirov-1}S. Cahangirov and S. Ciraci, Phys. Rev. B {\bf79}, 165118 (2009).
\bibitem{Keqiu-1}F. Ning, L.-M. Tang, Y. Zhang, and K.-Q. Chen, J. Appl. Phys. {\bf114}, 224304 (2013).
\bibitem{Keqiu-2}F. Ning, D. Wang, L.-M. Tang, Y. Zhang, and K.-Q. Chen, J. Appl. Phys. {\bf116}, 094308 (2014).
\bibitem{Keqiu-3}F. Ning, L. M. Tang, Y. Zhang, and K. Q. Chen, Sci. Rep. {\bf5}, 10813 (2015).
\bibitem{Lassen-1}B. Lassen, M. Willatzen, R. Melnik, and L. C. Lew Yan Voon, J. Mater. Res. {\bf21}, 2927 (2006).
\bibitem{Kishore-1}V. V. Ravi Kishore, N. \v{C}ukari\'{c}, B. Partoens, M. Tadi\'{c}, and F. M. Peeters, J. Phys.: Condens. Matter {\bf24}, 135302 (2012).
\bibitem{Kishore-2}V. V. Ravi Kishore, B. Partoens, and F. M. Peeters, Phys. Rev. B {\bf86}, 165439 (2012).
\bibitem{WangLW-1}Lin-Wang Wang, Jeongnim Kim, and Alex Zunger, Phys. Rev. B {\bf59}, 5678 (1999).
\bibitem{Xu-3}M. P. Persson and H. Q. Xu, Appl. Phys. Lett. {\bf81}, 1309 (2002).
\bibitem{Xu-4}M. P. Persson and H. Q. Xu, Phys. Rev. B {\bf70}, 161310 (2004).
\bibitem{Xu-5}M. P. Persson and H. Q. Xu, Nano Lett. {\bf4}, 2409 (2004).
\bibitem{Xu-6}M. P. Persson and H. Q. Xu, Phys. Rev. B {\bf73}, 035328 (2006).
\bibitem{Xu-7}M. P. Persson and H. Q. Xu, Phys. Rev. B {\bf73}, 125346 (2006).
\bibitem{Xu-11}G. H. Liao, N. Luo, Z. H. Yang, K. Q. Chen, and H. Q. Xu, J. Appl. Phys. {\bf118}, 094308 (2015).
\bibitem{Niquet-1}Y. M. Niquet, A. Lherbier, N. H. Quang, M. V. Fern\'{a}ndez-Serra, X. Blase and C. Delerue, Phys. Rev. B {\bf 73}, 165319 (2006).
\bibitem{Niquet-2}Y. M. Niquet, Phys. Rev. B {\bf74}, 155304 (2006).
\bibitem{Lind-1}E. Lind, M. P. Persson, Y.-M. Niquet, and L.-E. Wernersson, IEEE Trans. Electron Dev. {\bf56}, 201 (2009).
\bibitem{Xu-1}H. Q. Xu and U. Lindefelt, Phys. Rev. B {\bf 41}, 5979 (1990).
\bibitem{Xu-2}H. Q. Xu, Phys. Rev. B {\bf46}, 1403 (1992).
\bibitem{Chadi-1}D. Chadi and M. L. Cohen, Phys. Status Solidi B {\bf68}, 405 (1975).
\bibitem{Vogl-1}P. Vogl, H. P. Hjalmarson, and J. D. Dow, J. Phys. Chem. Solids {\bf44}, 365 (1983).
\bibitem{Jancu-1}J.-M. Jancu, R. Scholz, F. Beltram, and F. Bassani, Phys. Rev. B {\bf 57}, 6493 (1998).
\bibitem{Boykin-1}Timothy B. Boykin, G. Klimeck, and F. Oyafuso, Phys. Rev. B {\bf 69}, 115201 (2004).
\bibitem{Klimech-1} G. Klimech, R. C. Bowen, T. B. Boykin, and T. A. Cwik, Superlatt. Microstr. {\bf 27}, 519 (2000).
\bibitem{Carlo-1}A. Di Carlo, Semicond. Sci. Technol. {\bf 18}, R1 (2003).
\bibitem{Wolf-1}M. Wolfsberg and L. Helmholz. J. Chem. Phys. {\bf20}, 837 (1952).
\bibitem{Golub-1}G. H. Golub and C. F. Van Loan, {\em Matrix Computations}, 3rd ed. (The Johns Hopkins University Press, Baltimore, 1996).
\bibitem{Tinkham-1} M. Tinkham, {\em Group Theory and Quantum Mechanics}, (Dover Publications, INC., Mineola, New York, 2003).
\bibitem{Melvin-1} M. Lax, {\em Symmetry Principles in Solid State and Molecular Physics}, (Dover Publications, INC., Mineola, New York, 2001).
\bibitem{Yu-1} P. Y. Yu and M. Cardona, {\em Fundamentals of Semiconductors}, 4rd ed. (Springer, Heidlberg, Dordrecht, London, New York, 2010).

\end{thebibliography}
\end {document}